\documentclass[useAMS,usenatbib]{mn2e}
\usepackage{graphicx,longtable,color}
\usepackage{mathrsfs}
\usepackage{epsfig}
\usepackage{natbib}
\usepackage{color}
\usepackage{float}
\usepackage{amsmath}
\usepackage{times}
\usepackage{upgreek}
\usepackage[varg]{txfonts}
\usepackage{enumerate}
\usepackage{verbatim}
\usepackage{booktabs}
\usepackage{multirow}
\usepackage{amssymb}

\title[]{Radial alignment of elliptical galaxies by the tidal force of a cluster of galaxies}
\author[Yu Rong, Shu-Xu Yi, Shuang-Nan Zhang, Hong Tu]{Yu Rong$^{1}$\thanks{E-mail: rongyu@ihep.ac.cn}, Shu-Xu Yi$^{1}$, Shuang-Nan Zhang$^{1,2}$, Hong Tu$^{3}$\\
$^{1}$Key Laboratory of Particle Astrophysics, Institute of High Energy Physics, Chinese
Academy of Sciences, Beijing, China\\
$^{2}$National Astronomical Observatories,
Chinese Academy Of Sciences, Beijing, China\\
$^{3}$Physics Department \& Key Lab for Astrophysics, Shanghai Normal University, Shanghai, China}

\begin{document}
\maketitle
\begin{abstract}
Unlike the random radial orientation distribution of field elliptical galaxies, galaxies in a cluster are expected to point preferentially towards the center of the cluster, as a result of the cluster's
tidal force on its member galaxies. In this work an analytic model is formulated to simulate this effect.
The deformation time scale of a galaxy in a cluster is usually much shorter than the time scale of change of the tidal force; the dynamical process of the tidal interaction within the galaxy can thus be ignored. An
equilibrium shape of a galaxy is then assumed to be the surface of equipotential, which is the sum of the
self-gravitational potential of the galaxy and the tidal potential of the cluster at this location. We use a
Monte-Carlo method to calculate the radial orientation distribution of these galaxies, by assuming the NFW mass
profile of the cluster and the initial ellipticity of field galaxies. The radial angles show a single peak
distribution centered at zero. The Monte-Carlo simulations also show that a shift of the reference center from
the real cluster center weakens the anisotropy of the radial angle distribution. Therefore, the expected radial
alignment cannot be revealed if the distribution of spatial position angle is used instead of that of radial
angle. The observed radial orientations of elliptical galaxies in cluster Abell~2744 are consistent with
the simulated distribution.
\end{abstract}
\section{Introduction}
The total mass of a cluster of galaxies is dominated by dark matter, which couples only through gravity with
ordinary matter. Therefore, to map the density profile of a cluster, one needs to find how luminous matter in a
cluster is related to its gravitational potential. Each way of doing so leads to a method of cluster mass
mapping. Assuming some sort of dynamical balance established, one can model how the spatial distribution and
velocity dispersion of galaxies respond to its gravitational potential and then develop a technique to determine
the mass distribution \citep{1997ApJ...485L..13C,1982AJ.....87..945K}. Hydrostatic equilibrium supposed, the
density and temperature profiles of hot gas are related to the shape of the gravitational potential well
\citep{1980ApJ...241..552F,2002ApJ...576..152X}. Finding the manner of light rays bent by gravity gives birth to
the method of gravitational lensing \citep{1993ApJ...404..441K,2008ApJ...684..177U}. Although many approaches
have been made, the results from those are not always in good agreement
\citep{1998MNRAS.301..861W,1996MNRAS.283..431B}. Therefore, it is still useful to find other independent probes.

An extended object in a non-uniform gravitational field feels a distorting tidal force. The tidal force from the
mass of a cluster can change the shape of its galaxies in a way that the radial orientation angles (see the
upper panel of Fig.~\ref{fig:illu} for an illustration) of elliptical galaxies in a cluster show an anisotropic
distribution toward the cluster center different from the isotropic distribution of field galaxies; this
phenomenon is referred to as radial alignment (RA). If RA really exists and is caused by the tidal effect of the
cluster mass, then a new link between an observable, i.e. the radial angle distribution (RAD) of
elliptical galaxies, and the gravitational potential of the cluster can be found.

The idea that the cluster tidal force can cause RA was first proposed by Thompson \citep{1976ApJ...209...22T},
in order to explain ``a possible indication that the galaxies are preferentially aligned along the radius vector
to the center of the cluster" for the Coma cluster \citep{1975AJ.....80..477H}. In recent years, more
observational evidence has been found to support this idea
\citep{Pereira05,2006ApJ...644L..25A,2007ApJ...662L..71F}. Theoretically, \cite{1994MNRAS.270..390C} studied the dynamical
process of shape changing of an elliptical galaxy under the cluster's tidal field using N-body simulations, and
showed an alignment tendency between the galaxies' major axes with the radial direction.
\cite{2008ApJ...672..825P} found strong RA of dark matter substructures in a cluster halo in N-body simulations.
\cite{UsamiFuji} studied a gaseous ellipsoid of uniform density orbiting in the logarithmic potential, and
discussed the application of their result to the alignment of galaxies in rich clusters. They concluded that
inside a critical radius, galaxies would be disrupted by the tidal force of the cluster, whereas beyond the
critical radius the major axes of elliptical galaxies are locked in radial direction. None of the previous works
take the intrinsic ellipticity distribution of galaxies into consideration. In \cite{UsamiFuji}
all elliptical galaxies' intrinsic shape were set spherical, unlikely in real cases. In our work, the ellipticity distribution
of field elliptical galaxies, which can be thought unaffected by
environment, is used as the intrinsic ellipticity distribution.

This work aims to formulating an analytic model to simulate RA quantitatively. Section 1 is to
verify qualitatively that the shape changing timescale for most galaxies in a cluster is short enough compared
with the Keplerian motion period so that we can ignore the dynamical process of tidal distortion and assume an
equilibrium shape of the galaxy. In section 2 we calculate the equilibrium shape of an elliptical galaxy
analytically. Then the projection effects are included and RAD is generated by a Monte-Carlo method. Finally
the observed RADs of the clusters Abell~2744 are compared.
\section{Qualitative discussion}\label{sec:quality}
Consider a galaxy in a cluster, and the vector from the galaxy center to the cluster center is $L$. The enclosed mass
within radius $L$ is denoted by $M(L)$ and its potential field is $\Phi_{\rm{c}}$. The tidal force acting on a
mass element of the galaxy is approximately
\begin{equation}\label{eq:quality}
\ddot{r}_i\sim-\frac{\partial}{\partial {\hat{r}}_j}\frac{\partial \Phi_{\rm{c}}}{\partial {\hat{r}}_i}\mid_{\hat{r}=L}r_j,
\end{equation}
where $r$ and $\hat{r}$ are the vectors from the mass element to the galaxy center and to the center of the cluster, respectively. $i=1,2,3$ denote the three cartesian coordinate components of a vector. The Einstein summation convention is used in this work. We define a tidal potential corresponding to the tidal force, which is
\begin{equation}
\Phi_{\rm{T}}=\frac{\partial \Phi_{\rm{c}}}{\partial {\hat{r}}_i}\mid_{\hat{r}=L}r_i.
\label{t_p}
\end{equation}

\begin{figure}
\begin{center}
\includegraphics[width=0.48\textwidth]{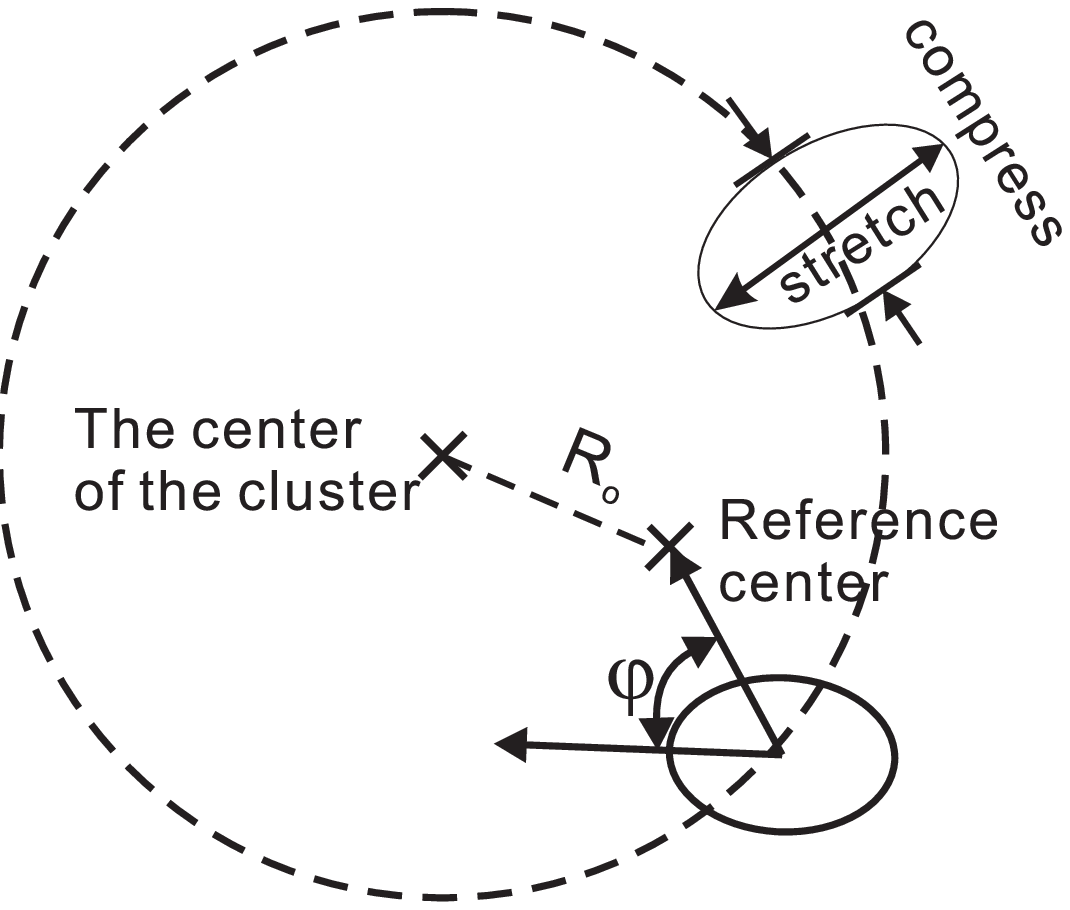}
\includegraphics[width=0.48\textwidth]{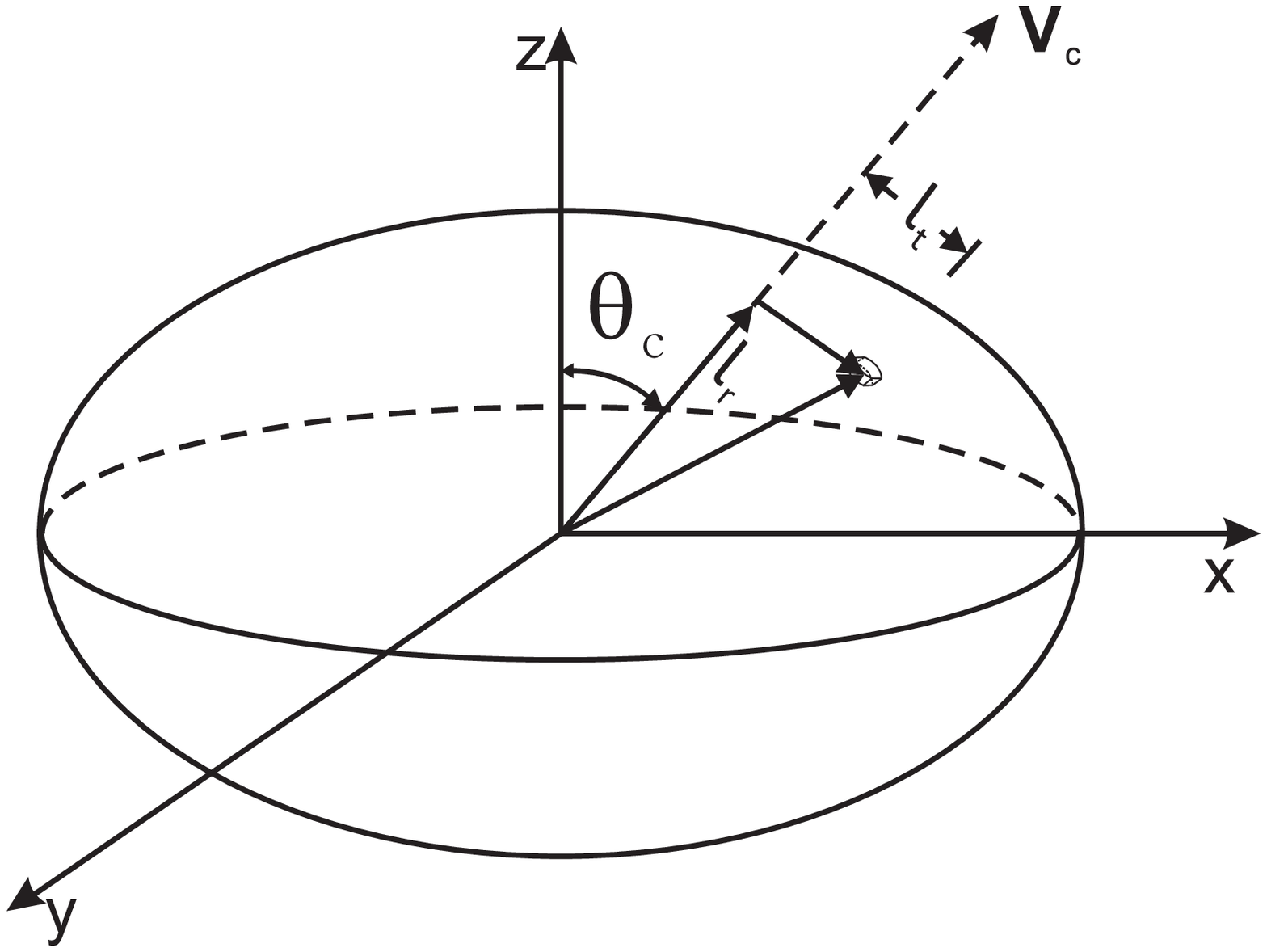}
\caption{\it{Upper panel}: \rm{}Illustration of radial orientation angle. $R_0=L_0/r_{\rm{s}}$ is the dimensionless offset distance, where $L_0$ is the distance between the real mass center of the cluster and the reference center used in observation, $\varphi$ is the angle between the direction of the projected elliptical galaxy's long axis and the vector pointing from the galaxy center to the reference center \it{Lower panel}:
\rm{}Definition of the coordinate system. $\vec{V}_c$ is the unit vector pointing to the direction of the center of cluster, with an angle $\theta_c$ between which and the $z$ axis. $l_t$ is the distance from the mass element to vector $\vec{V}_c$; $l_r$ is the projection of the mass element's position vector onto $\vec{V}_c$}
\label{fig:illu}
\end{center}
\end{figure}

The mass distribution of the cluster is treated as spherically symmetric, so its gravity acting on a mass element can be written as
\begin{equation}
\Phi_{\rm{c}}=-\frac{GM(\hat{r})}{|\hat{r}|}+\Phi_{\rm{M}}(\hat{r}),
\label{p_c}
\end{equation}
where $M(\hat{r})$ denotes the enclosed mass in radius $\hat{r}$, and $\Phi_{\rm{M}}(\hat{r})$ denotes the gravity potential caused by the mass gradient. Here since $-\nabla \Phi_{\rm{c}}=\frac{GM(\hat{r})}{|\hat{r}|^3}\hat{r}$ \citep{James87}, we obtain
\begin{equation}
\frac{G\cdot \nabla M(\hat{r})}{|\hat{r}|}-\nabla \Phi_{\rm{M}}(\hat{r})=0.
\label{p_c_l}
\end{equation}
Substituting Equations~(\ref{p_c}) and (\ref{p_c_l}) into Equation~(\ref{t_p}), and neglecting the variation of $M(\hat{r})$ in small scale, we obtain
\begin{equation}
\Phi_{\rm{T}}=\frac{GM(L)}{L^3}(\frac{1}{2}l_{\rm{t}}^2-l_{\rm{r}}^2),
\end{equation}
where $l_{\rm{r}}$ is the component of $r$ along $L$, and $l_{\rm{t}}$ is the perpendicular component, which are illustrated in Fig.~\ref{fig:illu}. Therefore the tidal force acting on an unit mass is obtained by $\ddot{r}_{\rm{T}}=-\nabla \Phi_{\rm{T}}$.

An unit mass element is also subjected to the self-gravity from the galaxy,
\begin{equation}
\begin{aligned}
\ddot{r}_{\rm{grav}}&=\ddot{r}_{\rm{e}}+\ddot{r}_{\rm{dm}}\\&\simeq\frac{GM_{\rm{G}}}{R_{\rm{G}}^2}+\frac{GM_{\rm{dm}}}{R_{\rm{G}}^2}\\&\simeq G\pi\rho_{\rm{G}}R_{\rm{G}}+\frac{GM_{\rm{dm}}}{R_{\rm{G}}^2},
\end{aligned}
\end{equation}
where $\ddot{r}_{\rm{e}}$ and $\ddot{r}_{\rm{dm}}$ are the gravity acting on the unit mass from the luminous and dark matter at the effective radius of a galaxy; $M_{\rm{G}}$ and $R_{\rm{G}}$ are the mass and effective radius of the luminous matter in the galaxy. The luminous matter is assumed homogeneous, with density $\rho_{\rm{G}}\sim 10^8M_{\odot}$. $M_{\rm{dm}}$ is the enclosed mass of the dark matter within the effective radius. In this work, we only consider the tidal distorsion within the effective radius. The density profile of the dark matter in an elliptical galaxy is \citep{Victor12}
\begin{equation}\label{equ:dm}
\rho_{\rm{dm}}=\frac{\rho_{\rm{c}}}{1+(r/r_{\rm{c}})^2},
\end{equation}
where $\rho_{\rm{c}}\sim 0.1M_{\odot}/\rm{kpc}^3$ is the central density, and $r_{\rm{c}}\sim 5\ \rm{kpc}$ is the scale radius. Therefore the dark mass enclosed within $r$ is $M_{\rm{dm}}=4\pi \rho_{\rm{c}}r_{\rm{c}}^3(\frac{r}{r_{\rm{c}}}-{\rm{arctan}}(\frac{r}{r_{\rm{c}}}))$. At the effective radius $R_{\rm{G}}\sim 10$~kpc, $\ddot{r}_{\rm{e}}\simeq G\pi \rho_{\rm{G}}R_{\rm{G}}\sim 10^9$, $\ddot{r}_{\rm{dm}}\simeq GM_{\rm{dm}}/R_{\rm{G}}^2\sim 10^8$, where $\pi G=1$ is used. Therefore $\ddot{r}_{\rm{dm}}\ll\ddot{r}_{\rm{e}}$, $\ddot{r}_{\rm{grav}}\simeq\ddot{r}_{\rm{e}}$, implying that the effect of the dark matter in an elliptical galaxy can be neglected within the effective radius. This has been comfirmed by \cite{Paolis95}, who found that dark matter inside the effective radius is negligible with respect to the luminous matter.

Define $\beta$ as the ratio of the tidal force and self-gravity,
\begin{equation}
\beta\equiv \frac{|\ddot{r}_{\rm{T}}|}{|\ddot{r}_{\rm{grav}}|}\simeq \frac{|\nabla \Phi_{\rm{T}}|}{|\ddot{r}_{\rm{grav}}|}\leq \frac{GM(L)/L^3}{G\pi \rho_{\rm{G}}R_{\rm{G}}}\frac{\partial (r^2)}{\partial r}\mid_{r=R_{\rm{G}}}\simeq\frac{M(L)}{\pi\rho_{\rm{G}}L^3},
\label{beta_value}
\end{equation}
which reflects the significance of tidal force compared with self-gravity. In the case of cluster A2744, the mass distribution can be modeled by a NFW profile with the parameter $r_{\rm{s}}\thickapprox160\ \rm{kpc}$ \citep{Zhang06}. We take $r_{200}$, i.e., the radius within which the mean density should be $200\times\rho_{\rm{c}}$, as the boundary of the cluster, and set $\rho_{\rm{G}}=10^8M_{\odot}/\rm{kpc}^3$, the same order of magnitude as the average density of
the Milky Way. We find that the galaxies with $\beta>1$ only compose 0.1\% the volume of the cluster. Thus for a rich cluster like A2744, the condition
$\beta\ll1$ is satisfied for most galaxies, which means that the tidal force is weak compared with the self-gravity of these galaxies.

\cite{1983A&A...121...97C} calculated the response of a star under the tidal force which is weak compared with its self-gravity. Analogous to a star, a galaxy will have a trace-free quadrupole oscillation, with a frequency $\omega_{\rm{osc}}\simeq \sqrt{\pi G\rho_{\rm{G}}}$. Therefore the deformation time scale of a galaxy should be $\tau_{\rm{D}}\simeq 1/\omega_{\rm{osc}}=1/\sqrt{G\pi \rho_{\rm{G}}}$. Equation~(\ref{beta_value}) is used to substituting $\pi \rho_{\rm{G}}$, we obtain $\tau_{\rm{D}}\simeq 1/\sqrt{GM(L)/\beta L^3}=\sqrt{\beta}\tau_{\rm{K}}$, where $\tau_{\rm{K}}\simeq 1/\sqrt{GM(L)/L^3}$ is the Keplerian time scale characterizing the rate of change of the tidal force. Since $\beta\ll 1$, we have $\tau_{\rm{D}}\ll \tau_{\rm{K}}$. It follows that the galaxy will remain close to the stationary equilibrium state determined by the instantaneous value of the tidal potential \citep{1983A&A...121...97C}, where the equilibrium shape of the galaxy will coincide with the iso-potential surface of the equilibrium state $\Phi_{\rm{eff}}=\Phi_{\rm{grav}}+\Phi_{\rm{T}}$.

\section{Quantitative calculation of the effect of tidal force on RAD of galaxies in a cluster}
\subsection{Coordinate system}\label{sec:cor}

Undoubtedly, an elliptical galaxy with a central luminous ellipsoid and dark matter halo component is a very complex entity, and therefore, we need to assume some necessary simplifications and assumptions in order to be able to study mathematically the deformation of such a complicated system under the tidal force from a cluster. Since we only consider the tidal distorsion within the effective radius $r\sim 10$~kpc, therefore hereafter the effect of dark matter is neglected (see section 2 for details). We assume that the surface of the luminous matter is an oblate ellipsoid \citep{Lambas92} at the effective radius, with ellipticity $\epsilon$ before tidal distortion, the equation of the shape can be written as,
\begin{equation}\label{equ:ellipsoid}
\frac{x^2+y^2}{\mu^2}+\frac{z^2}{\nu^2}=1,
\end{equation}
where $x$, $y$ and $z$ axes are the principal axes of the ellipsoid, and $\nu$ and $\mu$ are related by the
following equation,
\begin{equation}\label{equ:ellipticity}
\nu^2=(1-\epsilon^2)\mu^2.
\end{equation}
We require that the vector pointing to the center of the cluster falls in the plane $x-z$, thus the coordinate
system can be uniquely defined, as long as the center of the cluster is not on the $z$ axis.

The direction of the cluster center is denoted as:
\begin{equation}\label{equ:center}
\vec{V}_{\rm{c}}=(\sin\theta_{\rm{c}},0,\cos\theta_{\rm{c}}),
\end{equation}
where $\theta_{\rm{c}}$ is the angle between the direction of the cluster center and the $z$ axis. The
coordinate system and vector $\vec{V}_{\rm{c}}$ are illustrated in the lower panel of Fig.~\ref{fig:illu}.

\subsection{Deformation of a single elliptical galaxy}

In order to simplify the calculation, according to the homoeoid theorem \citep{James87}, we assume that the iso-potential surface of the gravity potential of the luminous matter at the effective radius coincides with the local surface of the luminous ellipsoid. Then the self-gravity potential of an elliptical galaxy at the effective radius before tidal distortion can be written as,
\begin{equation}\label{equ:potential}
\begin{aligned}
\Phi_{\rm{grav}}=A_1(x^2+y^2)+A_3z^2,
\end{aligned}
\end{equation}
where $A_1=(1-\epsilon^2)A_3$. It is easy to find $A_3\simeq G\pi\rho_{\rm{G}}$.

The contribution of tidal force is equivalent to adding a tidal potential $\Phi_{\rm{T}}$ upon the self-gravity
potential $\Phi_{\rm{grav}}$; here we take,
\begin{equation}\label{equ:eff}
\Phi_{\rm{eff}}=\Phi_{\rm{grav}}+\Phi_{\rm{T}}.
\end{equation}
A little calculation gives,
\begin{eqnarray}\label{tp}
\nonumber
\Phi_{\rm{T}}=-\frac{B}{4}((1-3\cos2\theta_{\rm{c}})x^2-2y^2\\
+(1+3\cos2\theta_{\rm{c}})z^2+6xz\sin2\theta_{\rm{c}}),
\end{eqnarray}
where $B=\frac{GM(L)}{L^3}$. Hence an explicit expression of $\Phi_{\rm{eff}}$
is given by:
\begin{equation}\label{equ:abc}
\Phi_{\rm{eff}}=ax^2+by^2+cz^2+dxy+exz+fyz,
\end{equation}
where $a=A_1-\frac{B}{4}(1-3\cos2\theta_c)$, $b=A_1+\frac{B}{2}$, $c=A_3-\frac{B}{4}(1+3\cos2\theta_c)$,
$e=-\frac{3B}{2}\sin2\theta_c$ and $d=f=0$. Therefore, the final shape of the distorted galaxy is fully described,
given the initial $\epsilon$, $\vec{V}_{\rm{c}}$, $\rho_{\rm{G}}$, together with the cluster mass profile
$M(L)$, which is taken as the NFW profile \citep{1996ApJ...462..563N},
\begin{equation}\label{equ:nfw}
M(r)=4\pi\rho_{\rm{0}}r^3_{\rm{s}}(\ln(1+r/r_{\rm{s}})-\frac{r/r_{\rm{s}}}{1+r/r_{\rm{s}}}),
\end{equation}
where $\rho_0$ and the ``scale radius" $r_s$ are parameters of the distribution. Define
$\delta_0\equiv\frac{\rho_0}{\rho_{\rm{c}}}$, where $\rho_{\rm{c}}$ is the critical density defined as
$\frac{3H^2}{8\pi G}$, where $H$ is the Hubble constant and $G$ is the gravitational constant,
\begin{equation}\label{equ:delta0andc}
\delta_0=\frac{200}{3}\frac{c^3}{\ln(1+c)-c/(1+c)},
\end{equation}
where $c$ is the so-called ``concentration parameter" defined as $c=\frac{r_{200}}{r_{\rm{s}}}$, where $r_{200}$
is the radius within which the mean density should be $200\times\rho_{\rm{c}} $ \citep{1996ApJ...462..563N}.
Substituting Equation~(\ref{equ:nfw}) into the expression of $B$, we get
\begin{equation}\label{equ:B}
B=4\pi G\rho_0r'^{-3}(\ln(1+r')-\frac{r'}{1+r'}),
\end{equation}
where $r'=r/r_{\rm{s}}$ is the dimensionless radius. Given $c$ and $r'$, $B$ can be obtained from Equations~(\ref{equ:delta0andc})
and (\ref{equ:B}). Analogously, $A_1$, $A_3$ and all the parameters from $a$ to $f$ of the ellipsoid can be
calculated.

\subsection{Projection effects}
From the view of an observer, the ellipsoidal shape of a galaxy is projected to become an ellipse. The next step
 is to determine the major axis of the projected ellipse and evaluate the angle between that and the
 direction toward the cluster center.
The observer is assumed in the direction $\vec{V}_{\rm{o}}$, which can be expressed as,
\begin{equation}\label{equ:obs}
\vec{V}_{\rm{o}}=(\sin\theta_{\rm{o}}\cos\phi_{\rm{o}},\sin\theta_{\rm{o}}\sin\phi_{\rm{o}},\cos\theta_{\rm{o}}).
\end{equation}

The problem is then reduced to such a constrained extremum value problem:
 To find the farthest point $\vec{X}$ from the vector $\vec{V}_{\rm{o}}$ in the surface of
\begin{equation}
\rm{Const}\it{}=ax^{\rm{2}}+by^{\rm{2}}+cz^{\rm{2}}+dxy+exz+fyz,
\end{equation}
 then the projection of $\vec{X}$ on the celestial sphere, $\vec{X}_{\rm{p}}$, is the direction of the major axis. Define
$\vec{V}_{\rm{cp}}$ as the projection of vector $\vec{V}_{\rm{c}}$ on the celestial sphere, we obtain the radial
orientation angle as,
\begin{equation}\label{equ:phi}
\varphi=\arccos\frac{\vec{V}_{\rm{cp}}\cdot\vec{X}_{\rm{p}}}{\mid\vec{V}_{\rm{cp}}\mid\mid\vec{X}_{\rm{p}}\mid}.
\end{equation}
 The details of solving the constrained extremum value problem are presented in the Appendix.

\subsection{RAD of elliptical galaxies}
In order to derive RAD of elliptical galaxies in a cluster, we use Monte-Carlo simulations. In each run, a
set of elliptical galaxies are generated, each of which is given a set of parameters sampled with certain
probability distributions, as discussed below.
\begin{figure}
\begin{center}
\includegraphics[width=0.49\textwidth]{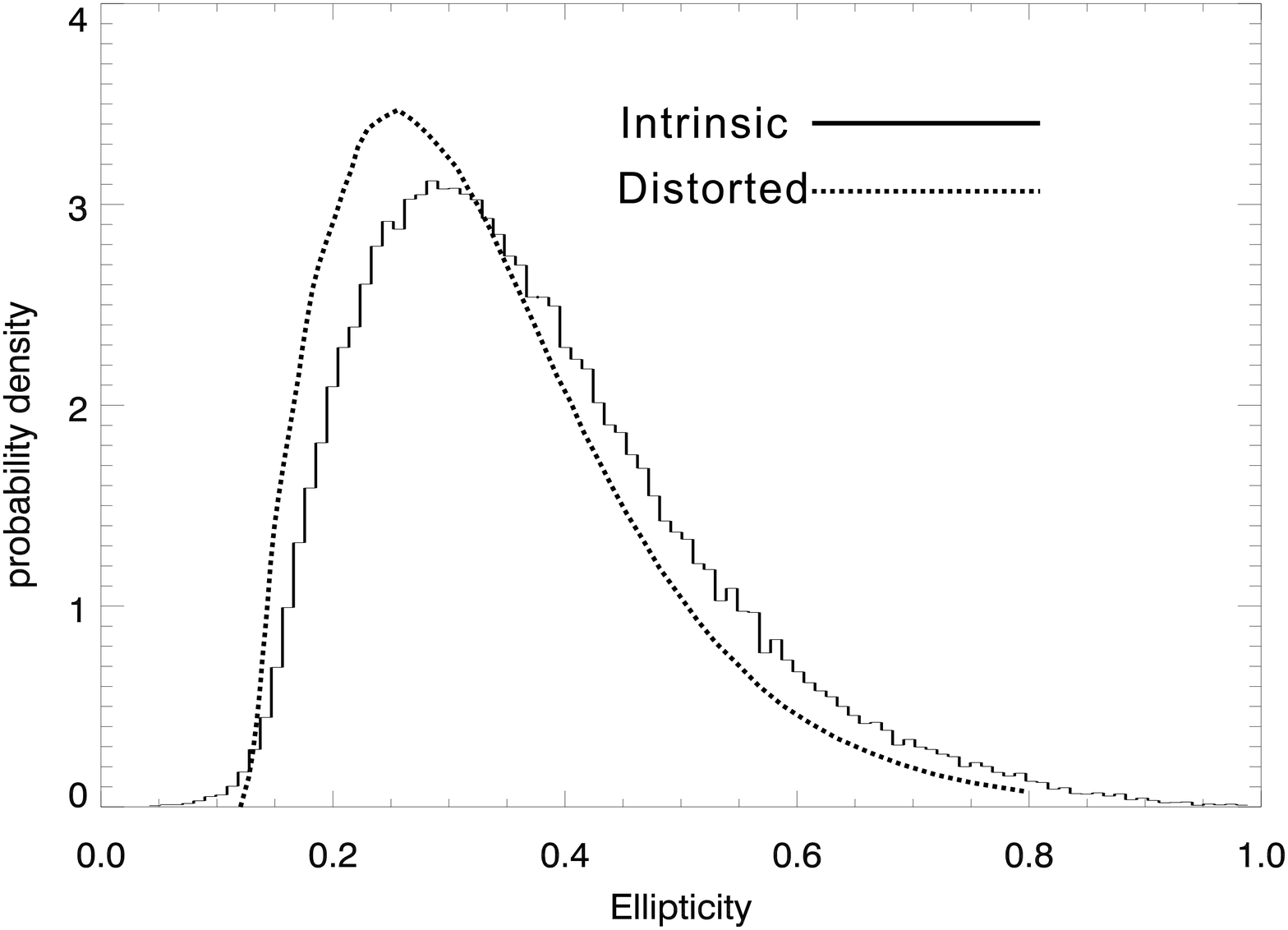}
\caption{The intrinsic (dashed line) and distorted (solid line) ellipticity distributions of galaxies.}
\label{fig:elp}
\end{center}
\end{figure}
With $M(r)$ given, each galaxy from the set results in a $\varphi_i$ ($i=1,2...N$). $c=10$ is used to simulate a cluster with high concentration like Abell~2744 (see section 4.1); note that $c=10$ is also the concentration of Coma cluster \citep{Rines03} where RA was found for the first time. The effective density of each galaxy is set to be $10^8$
times $\rho_{\rm{c}}$. In this run, 100,000
galaxies are generated, with an intrinsic ellipticity distribution in Fig.~\ref{fig:elp}, which is the
ellipticity distribution of field galaxies \citep{1991MNRAS.249..629F}. Here the ellipticity of a generated ellipsoid is defined as $\epsilon\equiv \frac{l_{\rm{max}}-l_{\rm{min}}}{l_{\rm{max}}}$, where $l_{\rm{max}}$ and $l_{\rm{min}}$ are the major and minor axes of the ellipsoid, respectively. The dimensionless projected distance of galaxies
are sampled with a probability inversely proportional to $r'^{2}_p$, from $r'_p=2.5$ to $r'_p=5$, corresponding to an
approximation to the projected galaxies number density distribution in \cite{Baier1976}. Under such
parameters, the maximum value of $\beta$ is 0.0057, which is far smaller than unity. For each galaxy, the direction toward the cluster center
$\vec{V}_{\rm{c}}$ and the direction toward the observer $\vec{V}_{\rm{o}}$ in the principal axes coordinate
system are sampled with an isotropic probability, corresponding to an isotropic position distribution, which is
assumed as the initial configuration of galaxies in the cluster.

\begin{figure}
\begin{center}
\includegraphics[width=0.49\textwidth]{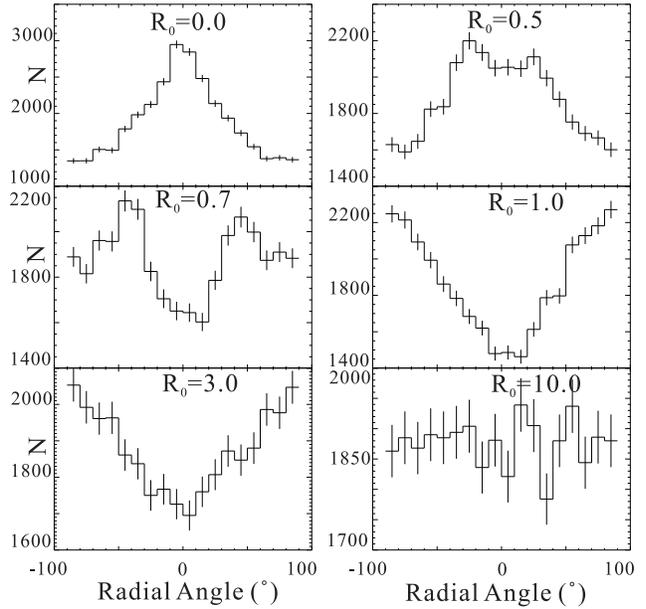}
\end{center} \caption{RADs for different center offset distances $R_0$.}\label{fig:shi}
\end{figure}

Fig.~\ref{fig:shi} shows RADs for the different center offset distances $R_0$, where $R_0\equiv L_0/r_{\rm{s}}$ is defined as the dimensionless offset distance, where $L_0$ is the distance between the real mass center of the cluster and the reference center used in observation. When $R_0=0$, the upper-left panel in Fig.~\ref{fig:shi} shows a single peak. Therefore if the reference center is the real cluster center, the major axes of the elliptical galaxies tend to point to the cluster center, i.e., we can see obvious radial alignment of the elliptical galaxies. However if the reference center has an offset distance from the real cluster center, since the major axes tend to point to the real center rather than the reference center, two peaks are found in RADs. As the center deviation distance increases, the two peaks depart from each other farther with increasing $R_0$. Then the two peaks merge at $-90/90$ degrees as $R_0$ keeps increasing. Finally, RAD becomes uniform distribution if $R_0\gg 0$, and at this moment, radial angle distribution becomes position angle distribution.

We also plot the distorted ellipticity distribution of ellipsoids after the simulation in Fig.~\ref{fig:elp}. The width of the ellipticity distribution is broadened about $\Delta w=w_2-w_1\simeq 0.025$, where $w_1$ and $w_2$ are the full widths at the half maximum (FWHM) of the ellipticity distributions before and after the simulation, respectively; meanwhile, the ellipticity distribution is shifted to higher values compared with the intrinsic one.

\begin{figure}
\begin{center}
\includegraphics[width=0.49\textwidth]{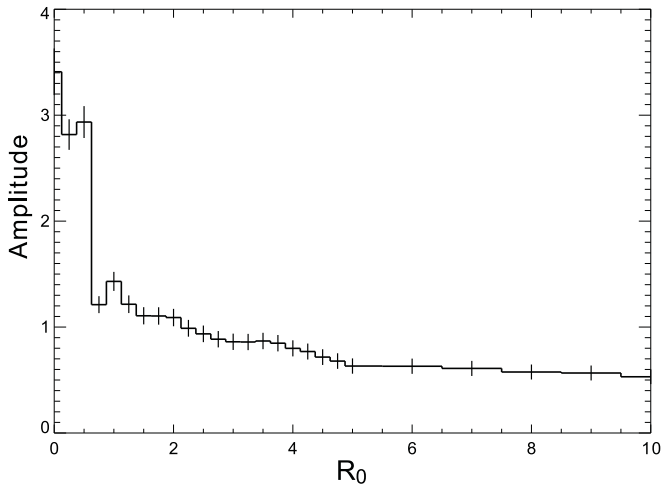}
\end{center} \caption{Amplitude of RAD versus $R_0$.}\label{fig:a_offset}
\end{figure}

In order to further reveal the effect when the center of the cluster is mis-identified observationally, we define the ``amplitude'' of RAD as $A_{\rm{rad}}=\frac{N_{\rm{max}}-N_{\rm{min}}}{\bar{N}}$, where $N_{\rm{max}}$, $N_{\rm{min}}$ and $\bar{N}$ are the highest point, lowest point and mean value of RAD. $A_{\rm{rad}}$ intuitively suggests the significance that a peak-type distribution deviates from uniform distribution, i.e., $A_{\rm{rad}}\simeq 0$ when the distribution is similar to uniform distribution, while $A_{\rm{rad}}\gg 0$ if there are significant peaks in the distribution. Fig.~\ref{fig:a_offset} shows $A_{\rm{rad}}$ as a function of the offset distance from the reference center to the true center of the cluster mass distribution. On the whole, $A_{\rm{rad}}$ decreases with increasing $R_0$. For $R_0\gg1$, a radial angle is equivalent to a position angle. The result in Fig.~\ref{fig:a_offset} explains why it is difficult to find RA with only the position angle distribution of elliptical galaxies in a cluster.

\section{Observed RAD of Abell~2744}
\subsection{Selection of Elliptical Galaxies}

Abell~2744 is one of the richest clusters of galaxies at intermediate redshift $z\sim 0.308$, known as a gravitational lens \citep{Merten11}. Its parameters are estimated as $r_{\rm{s}}=160\ \rm{kpc}$ \citep{Zhang06}, $r_{200}=2.506\ \rm{Mpc}$ \citep{Demarco03}, hence the
concentration parameter $c=r_{200}/r_{\rm{s}}\simeq 15.7$.
Such a high mass concentration is expected to show some signature in its RAD. We thus choose this cluster to
make the comparison with our theoretical work. The coordinate system is centered at $\alpha\ ({\rm{J2000.0}})=00^{\rm{h}}14^{\rm{m}}21.04^{\rm{s}}$, $\delta\ ({\rm{J2000.0}})=-30^{\circ}23^{\rm{m}}52.4^{\rm{s}}$ \citep{Boschin06}.
However, note that this cluster is an actively merging cluster \citep{Owers11}; therefore, the cluster central position cannot be determined unambiguously \citep{Zhang06}, which may bias RAD.

We retrieve HST (ACS/WFC) images in the F606W and F814W bands of two fields in the immediate vicinities of the cluster center (Date: 10/27/2009, PID: 11689, PI: Dupke). The associated images of A2744 are fully processed and drizzled. Photometry for objects is carried out using the Sextractor package \citep{SExtractor}. The configuration parameters are listed in Table~\ref{tcp}, and the magnitudes of the sources in the cluster are calibrated by the MAG\_ZEROPOINT parameter in the AB system. The derived source catalogs in the two bands are matched to obtain 3005 true objects. The parameter Class/Star$>0.9$ is used to get rid of stars, and then the removed stars are visually inspected in the images. The arcs and arclets, the spurious objects, the sources in the margins of images, and the sources inside other giant bright sources are visually identified and removed. The optical magnitudes of targets are corrected for the foreground extinction from the Galaxy according to the Schlegel-Finkbeiner-Davis Galactic reddening map \citep{Schlafly11}.

\begin{table}
\begin{tabular}{lc}
\hline
\textbf{Parameters} & Values\\
\hline
DETECT\_MINAREA&32\\
DETECT\_THRESH&1.3\\
DEBLEND\_NTHRESH&64\\
DEBLEND\_MINCONT&0.005\\
CLEAN\_PARAM&1.2\\
BACK\_SIZE&40\\
BACK\_FILTERSIZE&3\\
BACKPHOTO\_THICK&24\\
\hline
\end{tabular}
\caption{Sextractor configuration parameters for Abell~2744.}
\label{tcp}
\end{table}

\begin{figure}
\centering
\includegraphics[width=0.5\textwidth]{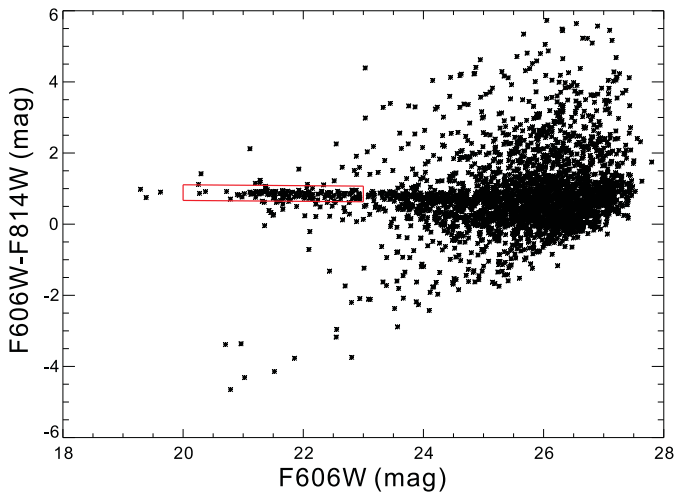}
\caption{CMD of Abell~2744. The galaxies with $color\sim -0.037m_{814}+1.610_{-0.2}^{+0.2}\ \rm{m}$ and $m_{814}<23.0\ \rm{m}$ are considered as the E/S0 galaxies and denoted by a box (red in the online article).}
\label{cmd}
\end{figure}

Limited by the ACS field size, the measurement reaches out up to a distance of $1.28\ \rm{Mpc}$ from the cluster center, which is about eight times of $r_{\rm{s}}$. These objects include both the foreground and background galaxies. In order to select the real cluster members, a color-magnitude diagram (CMD) is plotted in Fig.~\ref{cmd} using the magnitudes in the two filter bands, i.e., $color=m_{606}-m_{814}$, $mag=m_{606}$. Since the colors of elliptical galaxies are redder than spiral galaxies \citep{Baldry05}, and the Butcher-Oemler condition is satisfied, i.e., the blue galaxies are those at least 0.2 mag bluer than the cluster ridgeline \citep{Butcher84}, a linear fitting, i.e.,
\begin{equation}
color=k\cdot mag+d,
\label{linear_rs}
\end{equation}
where $k$, $d$ are the slope and intercept respectively, is applied to obtain the slope of the red sequence \citep{Brow92} in the CMD, which is known as the collection of E/S0 galaxies. Here $k\simeq -0.037, d\simeq 1.610$ are obtained; therefore we select the E/S0 galaxies in the color range of $color\sim -0.037m_{814}+1.610_{-0.2}^{+0.2}\ \rm{m}$. A faint end at about $m_{814}\sim 23.0\ \rm{m}$ is used to divide the member elliptical galaxies from the background galaxies \citep{Romano10}, which is approximately consistent with the rest-frame $r$-band absolute magnitude of $M_r\sim -17\ \rm{m}$. Finally 199 elliptical galaxies are obtained. These selected elliptical galaxies are then visually inspected in the images.

\subsection{RAD of Abell~2744}

\begin{figure}
\centering
\includegraphics[width=0.49\textwidth]{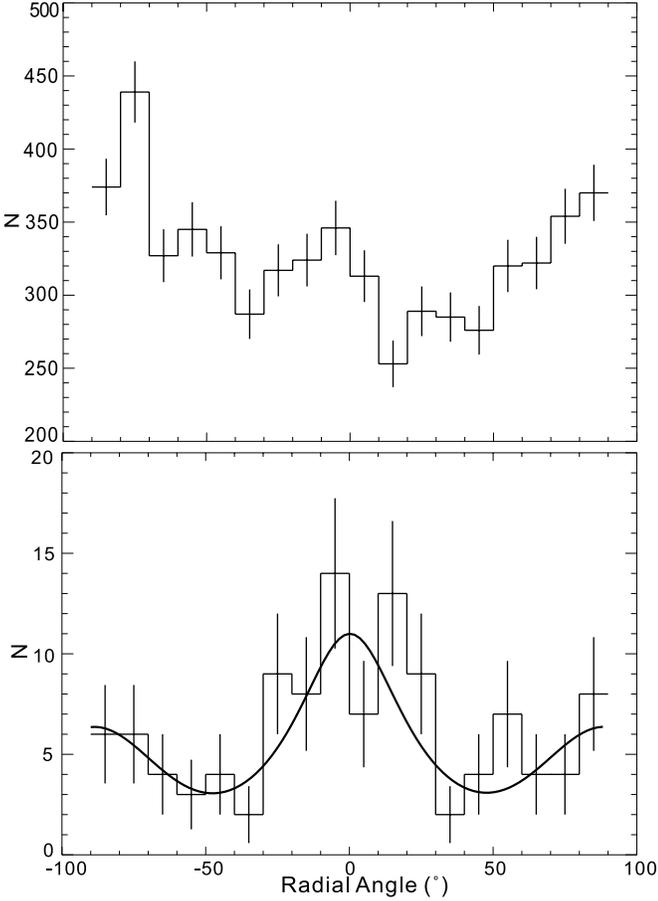}
\caption{Observed RADs of Abell~2744. The upper panel is RAD of the all sources; the bottom one is RAD of the elliptical galaxies. RAD of the elliptical galaxies is fitted by a double-Gaussian component plus a Lorentzian component.}\label{fig:observe}
\end{figure}

With SExtractor, we are able to quantitatively measure the orientations of the detected objects. We define the radial angle $\varphi=0$ when the major axis of a projected ellipsoid points to the cluster center, and $\varphi=\pm 90^{\circ}$ when the major axis orientates along the tangential direction. By separating the range of $-90^{\circ}\sim +90^{\circ}$ into 18 bins, we count the number of members in each bin. In order to obtain RAD with a high significance, we add together the number of galaxies in each bin of $\varphi$ from the two filter bands. RAD of the all sources of Abell~2744, including the foreground and background galaxies, and RAD of the elliptical galaxies in Abell~2744 are presented in Fig.~\ref{fig:observe}.

There are apparent tangential peaks centering at about $\pm 90^{\circ}$ in the upper panel of Fig.~\ref{fig:observe}, which should be produced by gravitational lensing \citep[e.g.][]{Smail97,Merten11}, since the shapes of background galaxies are stretched tangentially after lensing \citep{Joachim97}. Further more, since the selected elliptical galaxies are contaminated by some background galaxies, RAD in the bottom panel of Fig.~\ref{fig:observe} also show weak tangential peaks, which are fitted by a double-Gaussian component,
\begin{equation}
N=\frac{N_{\rm{G}}}{\sqrt{2\pi}\sigma_{\rm{G}}}e^{-(\varphi-\varphi_{\rm{G}})^2/2\sigma_{\rm{G}}^2}+\frac{N_{\rm{G}}}{\sqrt{2\pi}\sigma_{\rm{G}}}e^{-(\varphi-\varphi_{\rm{G}}-180^{\circ})^2/2\sigma_{\rm{G}}^2}.
\label{m_dgau}
\end{equation}
where $\varphi_{\rm{G}}=-90^{\circ}$ is fixed. Additionally, a peak is located at close to $\varphi=0$, which is fitted by a Lorentzian component,
\begin{equation}
N=\frac{N_{\rm{L}}\sigma_{\rm{L}}}{\pi[(\varphi-\varphi_{\rm{L}})^2+{\sigma_{\rm{L}}}^2]}.
\label{lorent}
\end{equation}
As a comparison, we also fit the central peak with a constant,
\begin{equation}
N=N_{\rm{C}}.
\label{m_linear}
\end{equation}
The two different fitting results are listed in Table~\ref{fit_two}. The relative error of $N_{\rm{L}}$ is $20\%$, suggesting that the central peak really exists, and thus the orientations of the elliptical galaxies are not random. $F$-test is performed to obtain the probability $p$ that the model DG+L is better than the model DG+C. Given the value of $F$, $F=\chi_1^2 \cdot dof_2/\chi_2^2 \cdot dof_1$, where $dof_1=13, dof_2=15$, and $\chi_1^2, \chi_2^2$ are the variances of the two models, respectively, thus we obtain $p\sim 83\%$.

\begin{table}\scriptsize
\begin{tabular}{@{}cccccccc@{}}
\hline
& & & DG+L & & & \\
 $N_{\rm{L}}$ & $\varphi_{\rm{L}}$ & $\sigma_{\rm{L}}$ & $N_{\rm{G}}$ & $\varphi_{\rm{G}}$ & $\sigma_{\rm{G}}$ & $\chi^2/dof$\\
 $8.3(1.7)\times10^2$\ & 0.17(4.11)\ & 24(7)\ & $3.0(1.0)\times10^2$\ & -90(fixed)\ & 22(8)\ & 14.0/13 \\
\hline
& & & DG+C & & & \\
  & $N_{\rm{C}}$ &  & $N_{\rm{G}}$ & $\varphi_{\rm{G}}$ & $\sigma_{\rm{G}}$ & $\chi^2/dof$\\
  & 4.5(0.6)\ &  & $5.3(5.4)\times10$\ & -90(fixed)\ & 6.8(9.5)\ & 27.8/15 \\
\hline
\end{tabular}
\caption{Fitting results of RAD of the elliptical galaxies. DG denotes a double-Gaussian component, L denotes a Lorentzian component, and C denotes a constant component. All errors are at $1\sigma$ throughout this work.}
\label{fit_two}
\end{table}

\section{Conclusion and Discussion}
We estimated qualitatively the shape changing time scale of a galaxy in a cluster, which is found to be much
shorter than its Keplerian time scale around the center of the cluster for most member galaxies. Thus we ignored
the dynamical process of shape changing and assumed an equilibrium shape of galaxies. The shape of an elliptical
galaxy under the tidal force of a cluster is calculated, by assuming that the density contour of the galaxy
coincides with the equipotential surface of the sum of the tidal potential and self-gravitational potential of
the galaxy. We then used the Monte-Carlo method to simulate RAD of galaxies in a cluster, taking the NFW
mass profile of the cluster and the initial ellipticity distribution as that of field galaxies. The Monte-Carlo simulations also find that the elliptical galaxies in a cluster tend to be radially aligned if the real physical center is chosen as the reference center. If the reference center has an offset distance from the real center, then the farther the reference point away from the cluster center, the weaker the anisotropy of RAD will be shown. Therefore in order to find significant RA, the reference center should be set to the real physical center of the cluster. In addition, the distribution of position angles which is equivalent to $R_0\gg 0$ cannot reflect RA. The observed RAD of cluster Abell~2744 is presented, using the data from HST. Comparing the observed RADs in
Fig.~\ref{fig:observe} with the simulated ones in Fig.~\ref{fig:shi}, we conclude that
 the observed non-uniform RADs are due to the tidal effects of these clusters.

In our model, the gravitational interaction between two nearby galaxies is neglected. We treated the galaxies as
if they were moving under a smooth averaged field. In fact when two galaxies get close, the tidal effects from
the nearby galaxy will weight over the entire cluster. This will increase the randomness of
 RAD of the whole cluster, because encounter events are random and the effects are local. The non-spherical symmetry and substructures of the
 cluster are also not included, since we applied the isotropic NFW mass profile. Those effects will be included in our following works.
 Our simulations place galaxies in the region where the self-gravity of the galaxies are far greater than the tidal force from the cluster, so tidal disruption of galaxies has not been taken into consideration.

In section 2, we assumed that a galaxy has a trace-free quadrupole oscillation under the weak tidal distorsion like a star \citep{1983A&A...121...97C} with a period $\tau_{\rm{osc}}\sim 0.1$~Gyr. However, an elliptical galaxy is a stellar system rather than a fluid system, and the oscillation can be quickly damped by Landau damping \citep{James87,Weinberg94}. Therefore, the time needed from the beginning of deformation to finally settling on the equilibrium state is indeed the damping time scale $\tau_{\rm{damp}}$, if $\tau_{\rm{damp}}>\tau_{\rm{osc}}$. In a finite stellar system like an elliptical galaxy, the time scale of damping the quadrupole oscillation \citep[$l=2$ mode]{Weinberg94} is shorter than $\tau_{\rm{osc}}/0.608\simeq 1.6\tau_{\rm{osc}}$ (see \cite{Weinberg94} for details). Therefore an elliptical galaxy will remain on the stationary equilibrium after $1.6\tau_{\rm{osc}}$, which is also shorter than $\tau_{\rm{K}}$. Therefore $\Phi_{\rm{eff}}=\Phi_{\rm{grav}}+\Phi_{\rm{T}}$ is a good approximation, where $\Phi_{\rm{T}}$ is instantaneous.

 In the Monte-Carlo simulations, the initial ellipticity distribution and the effective density of the galaxies are chosen as those of field galaxies.
 There is evidence showing that differences might exist between the density of cluster galaxies and field galaxies \citep{2011RAA....11..909P}; it is however possible that these
 differences are caused by the tidal effect we calculated here, which may change the density and ellipticity distributions of the member galaxies of a cluster.
  Although different initial ellipticity distribution
 and effective density chosen do not change our conclusion qualitatively, these will become important when
 analyzing RAD quantitatively. The effects of weak lensing are also not taken into consideration,
 which would cause images of some of galaxies in the far and back side of the cluster stretched along the tangential
 direction, and thus may weaken the observed RA by the tidal force of the cluster.
 
Finally, in this work we aim to study the possibility of radial alignment of elliptical galaxies, the original orientations of which are random, under the tidal force in a cluster. There are certainly other possibilities of radial alignment; for example, the galaxies entering the cluster could be radially pre-aligned, since tidal forces are ubiquitous. However, the issue is beyond the scope of the present work, and could be an interesting topic of future investigations. It is possible to even infer the original orientation and ellipticity distributions of galaxies by studying the observed distributions using our analysis method, if the gravitational field of the cluster is determined independently with other methods. Once again, this could be an interesting topic of future investigations.

\section*{Acknowledgments}
We appreciate the very insightful and constructive review report by the anonymous referee. We thank Professor Zu-Hui Fan for her helps on galactic dynamics. Drs. Yuan Liu and Jian Hu are thanked for their constructive suggestions and discussions. SNZ acknowledges partial funding support by 973 Program of China under grant 2014CB845802, by the National Natural Science Foundation of China under grant Nos. 11133002 and 11373036, and by the Qianren start-up grant 292012312D1117210, and by the Strategic Priority Research Program "The Emergence of Cosmological Structures" of the Chinese Academy of Sciences, Grant No. XDB09000000. Part of
this project is done under the support of the National Natural Science Foundation
of China Nos. 10878003, 10778752, 11003013, Shanghai Foundation No. 07dz22020,
and the Leading Academic Discipline Project of Shanghai Normal University
(08DZL805).

\bibliographystyle{mn2e}


\appendix

\section{Determining the major axis of an ellipsoid projected onto the celestial plane}
The projection of an ellipsoid is an ellipse. The problem is how to determine the major axis of the ellipse,
given the equation of the ellipsoid and the direction along which the ellipsoid is projected.

An ellipsoid described by equation,
\begin{equation}\label{equ:elpsoid}
ax^2+by^2+cz^2+dxy+exz+fyz=1,
\end{equation}
can be reexpressed in a matrix form,
\begin{equation}\label{equ:matrixform}
\vec{X}^{\rm{T}}A\vec{X}-1=0,
\end{equation}
where $\vec{X}\equiv(x,y,z)$ and $ A=\left(\begin{array}{ccc}
                                               a & d/2 & e/2 \\
                                               d/2 & b & f/2 \\
                                               e/2 & f/2 & c
                                             \end{array}\right)
$.
\\Suppose the direction of the observer is
\begin{equation}
\vec{n}=(\sin\theta\cos\phi,\sin\theta\sin\phi,\cos\theta),
\end{equation}
 so the problem is to find a point on the
ellipsoid farthest from the vector $\vec{n}$.

Define $\vec{v}_{\rm{b}}\equiv \lambda\vec{n}+\rho\vec{v}$, which is a bunch of rays that are parallel to
$\vec{n}$, where
 $\rho$ and $\lambda$ are real numbers and $\vec{v}$ is any unit vector that is perpendicular to
 $\vec{n}$. $\vec{v}$ can be written as,
\begin{equation}\label{equ:v}
\vec{v}=\frac{(-\cos\theta+\sin\theta\sin\phi j,-\sin\theta\cos\phi j,\sin\theta\cos\phi)}{\sqrt{j^2
\sin^2\theta-2j\sin\theta\cos\theta\sin\phi+\cos^2\theta+\cos^2\phi\sin^2\theta}},
\end{equation}
where $j\in(-\infty,+\infty)$.

The intersection points of the ellipsoid and the rays bunch satisfy the ellipsoid equation,
\begin{equation}\label{equ:vecAvec}
\vec{v}_{\rm{b}}^{\rm{T}}A\vec{v}_{\rm{b}}-1=0.
\end{equation}

Taken the expression of $\vec{v}_{\rm{b}}$, Equation~(\ref{equ:vecAvec}) can be written as,
\begin{equation}\label{equ:ercihanshu}
\lambda^2\vec{n}^{\rm{T}}A\vec{n}+\rho^2\vec{v}^{\rm{T}}A\vec{v}+2\lambda\rho\vec{n}^{\rm{T}}A\vec{v}-1=0,
\end{equation}
which is a quadratic function of $\lambda$. The ray $\vec{v}_{\rm{b}}$ is tangential to the ellipsoid when
Equation~(\ref{equ:ercihanshu}) has multiple roots, i.e.,
\begin{equation}\label{equ:banbieshi}
\rho^2=(\vec{v}^{\rm{T}}A\vec{v})^{-1}.
\end{equation}
The next step is to find $j$ that maximizes $\rho$. Using the expression of $A$ and $\vec{v}$,
Equation~(\ref{equ:banbieshi}) is rewritten as,
\begin{equation}\label{equ:rho}
\rho^2\propto\frac{j^2+\hat{a}j+\hat{b}}{j^2+\tilde{a}j+\tilde{b}},
\end{equation}
where
\begin{equation}\label{equ:aabb}
\begin{array}{l}
\hat{a}=-2\cot\theta\sin\phi\\
\hat{b}=\cos^2\phi+\cot^2\theta\\
\tilde{a}=\frac{U_{\rm{a}}}{D}\\
\tilde{b}=\frac{U_{\rm{b}}}{D}\\
\end{array}
\end{equation}
and $U_{\rm{a}}$, $U_{\rm{b}}$ and $D$ above can also be expressed explicitly in term of
$a,b,c,d,e,f,\theta,\phi$, in a lengthy but trivial way.

The value of $j$ that makes $\rho^2$ take an extreme value in Equation~(\ref{equ:rho}) is given by,
\begin{equation}\label{equ:result}
j=\frac{\tilde{b}-\hat{b}\pm\sqrt{(\hat{b}-\tilde{b})^2-(\tilde{a}-\hat{a})(\hat{a}\tilde{b}-\tilde{a}\hat{b})}}{\hat{a}-\tilde{a}},
\end{equation}
Substitute $j$ of Equation~(\ref{equ:result}) into Equation~(\ref{equ:v}), we can get both the major and minor
axes of the projected ellipse.
\end{document}